\documentclass[12pt,draftclsnofoot,onecolumn]{IEEEtran}
\usepackage{blindtext, graphicx,lipsum}
\usepackage[dvipsnames]{xcolor}
\usepackage{cuted,float}

\usepackage{bigints}
\usepackage{caption}
\usepackage{subcaption}
\newlength\figureheight 
\newlength\figurewidth 
\usepackage{cite}
\usepackage{url}
\usepackage{blindtext, graphicx}
\usepackage{amsmath,amssymb,mathtools,bm,etoolbox}

\usepackage{stackengine}

\usepackage{suffix}
\usepackage{stfloats}

\usepackage{tkz-euclide}
\usepackage{multicol}
\usepackage{tikz}
  \newlength\fheight
\newlength\fwidth
\usepackage{tkz-euclide}
\usepackage[utf8]{inputenc}
\usepackage{pgfplots} 
\usepackage{pgfgantt}
\usepackage{pdflscape}
\pgfplotsset{compat=newest} 
\pgfplotsset{plot coordinates/math parser=false}
\pgfplotsset{every  tick/.style={black,},ylabel style={font=\tiny},xlabel style={font=\tiny},tick label style={font=\tiny},legend style= {font=\scriptsize},
minor x tick num=1,minor y tick num=1,xminorticks=true,yminorticks=true,}
\DeclareMathOperator*{\argmax}{arg\,max}
\DeclareMathOperator*{\argmin}{arg\,min}

\def\endthebibliography{%
  \def\@noitemerr{\@latex@warning{Empty `thebibliography' environment}}%
  \endlist
}
\usepackage{amsthm}
\usetikzlibrary{arrows,%
                plotmarks}
\usepackage{tikz-3dplot}
\usepackage{balance}
\newtheorem*{remark}{Remark}

\usepackage{algorithm,algorithmic}
\usepackage{caption}
\begin{document}

\title{Joint Beamforming and Interference Cancellation in MmWave Wideband Full-Duplex Systems}

\author{Elyes~Balti,~\IEEEmembership{Student~Member,~IEEE,}
        and~Brian~L.~Evans,~\IEEEmembership{Fellow,~IEEE}
\thanks{Elyes Balti and Brian L. Evans are with the Wireless Networking and Communications Group, Department of Electrical and Computer Engineering, The University
of Texas at Austin, Austin, TX 78712 USA (e-mail: ebalti@utexas.edu, bevans@ece.utexas.edu).}
}

\maketitle

\begin{abstract}
Full-duplex (FD) systems have the capability to transmit and receive at the same time in the same frequency band.
FD systems can reduce congestion and latency and improve coverage and spectral efficiency.
As a relay, they can increase range and decrease outages.
Full-duplex (FD) wireless systems have been emerging as a practical solution to provide high bandwidth, low latency, and big data processing in millimeter wave and Terahertz systems to support cellular networks, autonomous driving, platooning, advanced driving assistance and other systems. 
However, FD systems suffer from loopback self-interference that can swamp the analog-to-digital converters (ADCs) resulting in very low spectral efficiency. In this context, we consider a cellular system wherein uplink and downlink users independently communicate with FD base station. The proposed contributions are (1) three hybrid beamforming algorithms to cancel self-interference and increase the received power, and (2) evaluation of outage probability, spectral efficiency, and energy efficiency of the proposed algorithms. We consider full-digital beamforming and upper bound as benchmarks. Finally, we show the resiliency of Algorithm 2 against self-interference in comparison with Algorithms 1 and 3, as well as conventional approaches such as beam steering, angle search and singular value decomposition.
\end{abstract}
\IEEEpeerreviewmaketitle

\section{Introduction}
With the increase in demand for data rates, cellular networks operating below 7 GHz have been unable to satisfy the growing number of human and machine subscribers due to bandwidth scarcity and expensive access licenses. In this context, using millimeter wave (mmWave) bands, which refers to the frequency band from 10 to 300 GHz,\footnote[1]{Although a rigorous definition of mmWave frequencies would place them between 30 and 300 GHz, industry has loosely defined them to include the spectrum from 10 to 300 GHz.} has been adopted in several standards to address spectrum scarcity \cite{j4,e1} such as IEEE 802.11ad and 802.11ay Wi-Fi standards and 5G New Radio (NR) in 3GPP Release 15 \cite{e2,e3,release15}.

MmWave technology has gained enormous attention both in academia and industry not only because it provides a link budget of several Gbps of data rate, but also it is a way to support ultra-dense cellular networks. In addition, the performance of the mmWave technology can be further enhanced when considering full-duplex (FD) systems. Such systems have already attracted interest in 5G networks because FD bidirectional links double the rate compared to a classical half-duplex (HD) relay \cite{j3}. Because of these advantages, FD systems can be a potential candidate for mmWave applications where large bandwidth and high spectral efficiency are required for big data processing. For example, vehicular-to-everything (V2X) applications such as platooning require low latency offered by FD relaying \cite{v2xfd}. In addition, FD is currently considered in 3GPP Release 17 for a mmWave integrated access and backhaul solution \cite{release17,5GFD21}.

Since FD systems transmit and receive at the same time and in the same frequency band, FD systems are exposed to self-interference (SI) which substantially degrades spectral efficiency \cite{zf}. The main challenge of mmWave FD systems is how to design robust precoders and combiners to cancel the interference and make the FD operation feasible \cite{chal1,chal2}. \textcolor{black}{The authors in \cite{zfjournal} formulated the optimization problem and applied the alternating projection method between the Zero-Forcing null-space and the subspace of the constant amplitude constraint. This method results in resilient hybrid beamformers design to combat the SI and maximize the sum spectral efficiency}. Table \ref{enab} illustrates the amount of SI that needs to be suppressed to enable FD operation for the different network generations.

\begin{table*}[t]
\renewcommand{\arraystretch}{1}
\caption{Amount of Self-Interference Wiped Out to Enable Full-Duplex Operation \cite{s6,s16,s37,s42,s43,s46,s47,s48,surveyfd}.}
\label{enab}
\centering
\begin{tabular}{cccccc}
\bfseries Generation & \bfseries Technology/Medium Access & \bfseries \bfseries Channel Bandwidth & \bfseries Transmit Power & \bfseries Noise Power & \bfseries SI Cancellation\\
\hline
1G & AMPS/FDMA & 30 KHz & up to 60 dBm & -129 dBm & 189 dB\\
\hline
2G & GSM/TDMA& 200 KHz & 36 dBm & -121 dBm & 157 dB\\
\hline
3G & WCDMA/UMTS & 5 MHz & 43 dBm & -107 dBm & 150 dB\\
~ & CDMA 2000 & 1.25 MHz & 43 dBm & -113 dBm & 156 dB\\
\hline
4G/LTE & LTE-Advanced&20 MHz & 46 dBm & -101 dBm & 147 dB\\
~&(OFDMA/SC-FDMA)&~&~&~&~\\
~ & WIMAX/ Scalable OFDMA & 10 MHz & 43 dBm & -104 dBm& 150 dB\\
\hline
5G & BDMA/ FBMC & 60 GHz & 20 dBm & -96 dBm & 116 dB\\
~ & 802.11ac - Gigabit Wi-Fi & 20, 40, 80, 160 MHz& 20 dBm& -91 dBm& 112 dB\\
~&(taunted as 5G Wi-Fi)&~&~&~&~\\
~&802.11ad - Wireless Gigabit & 2 GHz& 20 dBm & -81 dBm & 101 dB\\
~&(Microwave Wi-Fi)&~&~&~&~\\
~& 802.11af - White-Fi & 5, 10, 20, 40 MHz& 20 dBm&-98 dBm &118 dB\\
\hline
\end{tabular}
\\
\rule{0in}{1.2em}$^\dag$ \scriptsize{Advanced Mobile Phone Service (AMPS), Frequency Division Multiple Multiple Access (FDMA), Global Systems for Mobile Communications (GSM), Time Division Multiple Access (TDMA), Code Division Multiple Access (CDMA), Wideband CDMA (WCDMA), Universal Mobile Telecommunications Systems (UMTS), Long Term Evolution (LTE), Orthogonal/Single Carrier Frequency Division Multiple Access (OFDMA/SC-FDMA), Worldwide Interoperability for Microwave Access (WIMAX), Beam Division Multiple Access (BDMA), Filter Bank Multi-Carrier (FBMC)}.
\end{table*}

\subsection{Taxonomy of SI Cancellation Techniques}
Passive and active methods to cancel SI are surveyed next.

\subsubsection{Passive Suppression}
Passive SI suppression is based on separating the transmit and receive RF chains. Passive cancellation techniques rely on antenna directivity combined with physical separation of the antennas, polarization, and use of additional RF absorbing materials \cite{s37,s61}. When each of these techniques is carried out as standalone solution or in combination with other passive techniques, the primary objective is to isolate the transmit and receive RF chains as much as possible. Below, we present the passive SI cancellation approaches available in the literature with the relative advantages and drawbacks as well as their efficiencies in canceling the SI.

\textit{Antenna directionality} has been proposed as a passive technique to cancel the SI since it is easy to implement, it provides directional diversity, and it is suitable to narrowband scenarios. Although this technique can achieve about 30 dB of SI reduction, it is not suitable for wideband systems due to the large range of wavelengths needed to support the larger bandwidth \cite{s66,s74}.

\textit{Antenna placement} can be more efficient than antenna directionality as it is robust in narrowband scenarios and can achieve about 47 dB of SI reduction \cite{s6,s14}. However, this technique suffers from severe amplitude mismatch and requires manual tuning; hence, it is not adaptive to the environment \cite{s47,s75}. In addition, the SI can be substantially mitigated by \textit{cross-polarization} which can suppress about 50 dB of SI. This technique can be applied to separate and shared antennas, and to small-factor devices with duplexers \cite{s7,s19,s62,s74}. Table \ref{factor} summarizes the form factor dimensions of FD devices.
\begin{table}[b]
\renewcommand{\arraystretch}{1}
\caption{Reference Form Factor Dimensions for Full-Duplex Devices \cite{s68}.}
\label{factor}
\centering
\begin{tabular}{ccc}
\bfseries Full-Duplex & \bfseries Access Point & \bfseries \bfseries Form Factor\\
\bfseries Devices & \bfseries Type & \bfseries Dimension\\
\hline
Base Station & Femto  & 236 x 160 x 76 mm\\
~ & Pico & 426 x 336 x 128 mm\\
~ & TETRA  & 55 x 143 x 57 mm\\
\hline
User Equipment & Netbook& 285 x 202 x 27.4 mm\\
~& Tablet PC & 241.2 x 185.7 x 8.8 mm\\
~ & Smart Phone & 123.8 x 58.6 x 7.6 mm\\
~ & PDA & 132 x 66 x 23 mm\\ 
\end{tabular}
\end{table}
\subsubsection{Active Suppression}
Active suppression approaches use active components and leverage knowledge of a node's own SI signal in generating a cancellation signal to be subtracted from the received signal \cite{s26,s61,s70}. Active cancellation can be analog or digital \cite{s62}. Active cancellation applied before digitization of the received signal is termed \textit{active analog cancellation} whereas the active cancellation method employed to cancel the residual SI within the received signal after digitization is termed \textit{digital cancellation} \cite{s16,s40,s72,s73,s74}. 
Most active cancellation techniques are carried out in the active analog circuit domain. Below, we discuss the active analog and digital cancellation techniques along with their advantages and limitations.\\

\textbf{\em Analog Cancellation:} This approach aims to suppress the SI before the low-noise amplifier (LNA) and analog-to-digital converters (ADCs). The \textit{Balun circuit} is an analog circuit that reduces SI by about 45 dB. The baseline of this circuit is to generate an inverted version of the received signal for cancellation. In addition, this circuit is not limited in terms of bandwidth or power, and it can adapt to the environment without requiring manual tuning. The main drawback of this circuit is it incurs additional non-linearity from the noise canceling circuit, and hence the SI cancellation is not adequate \cite{s46,s60,s74}. Furthermore, an \textit{Electric Balance Duplexer} has been proposed as an SI analog canceler and highly depends on the frequency. This duplexer, which uses one antenna and hence it is cost-efficient, is suitable for small form-factor devices.  It is tunable over a wide frequency range and not constrained by the specific separation distance. However, this device is frequency dependent and requires manual tuning. In addition, this duplexer does not have good power handling capability and is prone to non-linear IB distortions \cite{s78,s79,s80}. For example, the \textit{QHx220 chip} is an analog circuit that suppresses about 45 dB of SI. This chip is beneficial in a way that provides extra RF chain; however, it is non-adaptive to the environment and difficult to implement for wideband systems \cite{s15,s16,s32,s74}.\\ 

\textbf{\em Digital Cancellation:}
Digital cancellation can be coupled with analog cancellation or go it alone. When coupled with an analog cancellation method, about 60 dB of SI reduction can be achieved because both SI and noise can be suppressed. However, it suffers from distortion due to non-ideality of transmitter and receiver components \cite{s7,s14}. Without analog cancellation, this mode can reduce only 10 dB of SI. Although the digital circuit, as the last line of defense, could eliminate the residual SI after cancellation, it is limited due to hardware impairments such as I/Q imbalance \cite{s6,s15,s16,s37}.

\subsection{Contributions}
In this paper, we consider a dual-hop FD base station (BS) independently communicating with uplink and downlink user equipment (UEs). We provide a channel model for the SI leakage to quantify the SI received power that affects the uplink UE. Our goal is to design robust beamformers to maintain the uplink user rate; the downlink user is immune to the interference.  The key contributions follow:
\begin{itemize}
    \item Present the system model wherein we discuss the signal model; uplink, downlink and SI channels; codebooks; and hardware structure.
    \item Present three algorithms for hybrid beamforming designs. The analog and digital stages are jointly designed by the first two algorithms while they are independently designed for the third.
    \item Analyze spectral efficiency, outage probability and energy efficiency and provide benchmarking in terms of upper bound and full-digital beamforming design to quantify the losses incurred by the SI for the proposed algorithms. 
    \item Compare robustness of each design algorithm against interference in the uplink and downlink. We also draw conclusions about tradeoffs in each algorithm.
\end{itemize}

\subsection{Structure}
The paper is organized as follows. Section II presents the system model wherein the channels models, array structure and codebooks are analyzed. The proposed beamforming designs are detailed in Section III while the performance analysis is reported in Section IV. Numerical results and concluding remarks are provided by Sections V and VI, respectively.

\subsection{Notation}
Bold lower and upper case letters represent vectors and matrices, respectively. $\mathbb{C}^{a\times b}$ denotes the space of complex matrices of dimensions $a\times b$. $(\cdot)^T$ and $(\cdot)^*$ represents the transpose and Hermitian, respectively. $\|\bold{X}\|_{F}$ is the Frobenius norm of matrix \textbf{X}. \text{Tr}(\textbf{X}) is the trace of matrix \textbf{X}, $\mathbb{E}[\cdot]$ is the expectation operator, and $\mathbb{P}[\cdot]$ is the probability measure. \text{det}(\textbf{X}) is the determinant of \textbf{X} and $\otimes$ is the Kronecker product.

\section{System Model}

\begin{figure}[H]
    \centering
    \includegraphics[width=\linewidth]{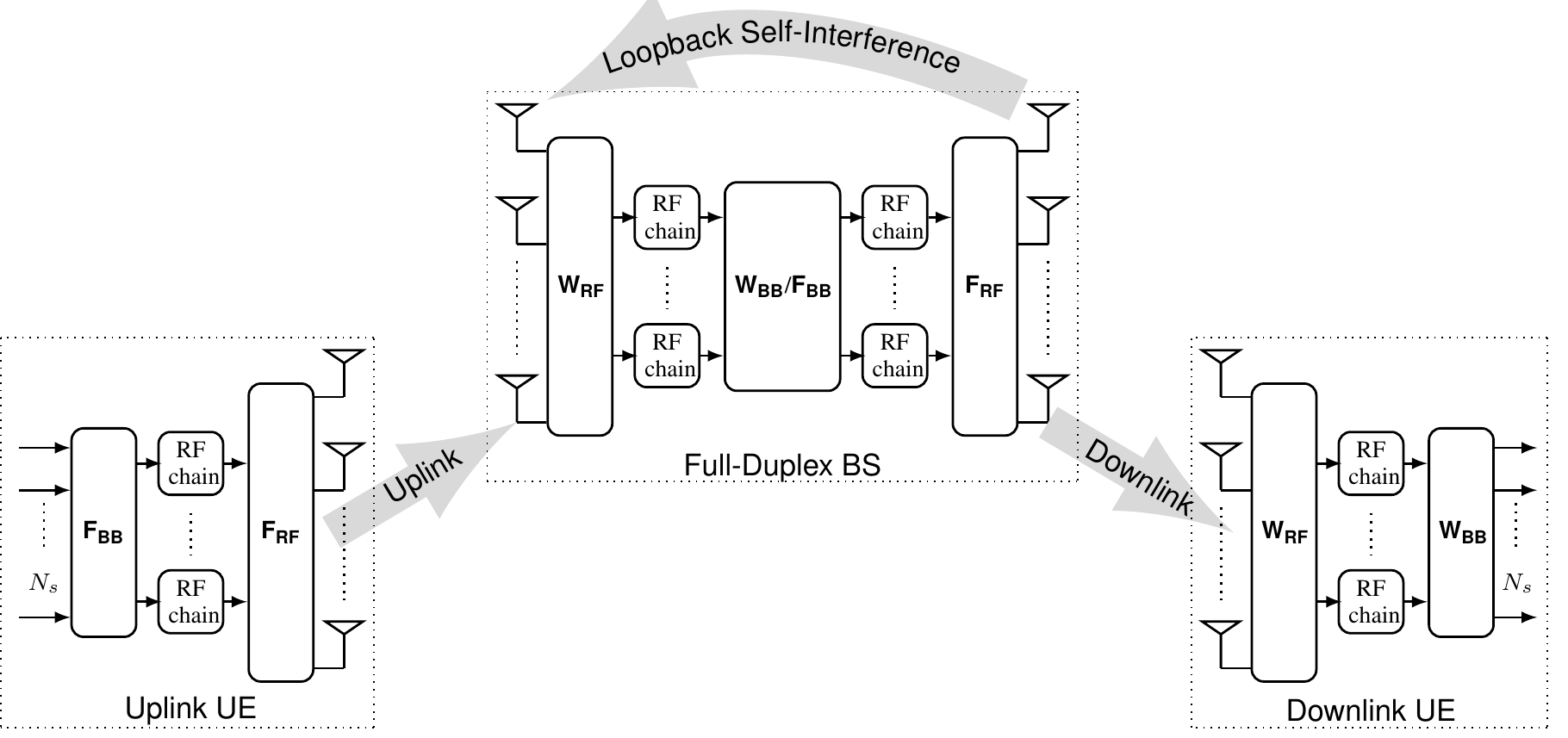}
    \caption{Hybrid architecture of dual-hop FD relay channel. The uplink UE sends the data to BS independently from the data intended to the downlink UE sent from BS. Since the BS transmits and receives simultaneously at the same frequency band, we model the SI leakage by the $n$-th channel tap $\textsf{H}_s[n]$.}
    \label{system}
\end{figure}
The proposed system uses OFDM signaling with $K$ subcarriers. At the $k$-th subcarrier, the symbols \textbf{s}[k] are transformed to the time domain using the $K$-point IDFT. The CP of length $L_c$ is then appended to the time domain samples before applying the precoder. The OFDM block is formed by the CP followed by the $K$ time domain samples. The data symbols follow $\mathbb{E}[\textbf{s}[k]\textbf{s}[k]^*]=\frac{\rho}{KN_s}\boldsymbol{I}$, where $\rho$ is the total average transmit power for the data per OFDM system, i.e.\ without considering the CP. We assume the maximum delay spread in the channel is within the cyclic prefix (CP) duration. Note that this system description applies equally to uplink and downlink transmission.

For uplink, the received signal at the BS and the $k$-th subcarrier is given by
\begin{equation}
\begin{split}
y_{\text{uplink}}[k] =& \underbrace{\sqrt{\rho_u}\textbf{W}^*_{\text{BS}}[k]\textbf{H}_u[k]\textbf{F}_{\text{UE}}[k]\textbf{s}_u[k]}_{\textsf{Desired Signal}}\\& + \underbrace{\sqrt{\rho_s} \textbf{W}^*_{\text{BS}}[k]\textbf{H}_s[k]\textbf{F}_{\text{BS}}[k]\textbf{s}_d[k]}_{\textsf{Self-Interference}} +  \underbrace{\textbf{W}^*_{\text{BS}}[k] \textbf{n}_{\text{BS}}[k]}_{\textsf{AWGN}}       
\end{split}
\end{equation}
where $\textbf{W}_{\text{BS}}[k],~\textbf{F}_{\text{BS}}[k]$, and $\textbf{F}_{\text{UE}}[k]$ are the $k$-th full-digital combiner, precoder at BS and $k$-th full-digital precoder at the uplink UE, respectively. $\textbf{H}_u[k]$ and $\textbf{H}_s[k]$ are the $k$-th uplink and SI subcarriers, respectively, while $\textbf{s}_u[k]$, $\textbf{s}_d[k]$ and $\textbf{n}_{\textbf{BS}}[k]$ are the UE data sent to BS, the BS data sent to downlink UE, and the additive white Gaussian noise (AWGN) at the BS, respectively. Note that $\rho_u$ and $\rho_s$ are the received power at BS and SI power, respectively.

For downlink scenario, the received signal at UE in the $k$-th subcarrier is expressed by
\begin{equation}
y_{\text{downlink}}[k] = \sqrt{\rho_d}\textbf{W}^*_{\text{UE}}[k]\textbf{H}_d[k]\textbf{F}_{\text{BS}}[k]\textbf{s}_d[k] + \textbf{W}^*_{\text{UE}}[k] \textbf{n}_{\text{UE}}[k]
\end{equation}
where $\textbf{W}_{\text{UE}}[k]$ is the $k$-th full-digital combiner at downlink UE, $\rho_d$ is the received power at UE, $\textbf{n}_{\text{UE}}[k]$ is the AWGN at the UE, and $\textbf{H}_d[k]$ is the downlink $k$-th subcarrier.

Unlike the downlink scenario, the uplink received signal is corrupted not only by the noise but also by the SI leakage occurred at the FD BS.
\subsection{Channel Model}
In this work, we assume that the MIMO channels for uplink and downlink are wideband, having a delay tap length $N_c$ in the time domain. The $n$-th delay tap of the channel is represented by a $N_{\text{R}} \times N_{\text{T}}$ matrix, $n = 1,\ldots,N_c-1$,  which, assuming a geometric clusters and rays based channel model given by \cite[Eq.~(6)]{anum}
\begin{equation}\label{channel}
\textsf{H}[n] = \gamma \sum_{c=0}^{C-1}\sum_{\ell=0}^{L-1} \alpha_{c,\ell} p_{\rm{rc}}(n T_s - \tau_{c,\ell}) \textbf{a}_r(\phi_{c,\ell}^r,\theta_{c,\ell}^r)   \textbf{a}_t^*(\phi_{c,\ell}^t,\theta_{c,\ell}^t)
\end{equation}
where $\gamma$ is a scaling factor to normalize the channel energy, $T_s$ is the signaling interval, $C$ is the number of clusers, $L$ is the number of rays per cluster, $\alpha_{c,\ell}$ is the complex gain of $\ell$-th ray in $c$-th cluster, $p_{\rm{rc}}(\tau)$ is the raised cosine filter for the pulse shaping evaluated at $\tau$, $\tau_{c,\ell}$ is the delay of $\ell$-th ray in $c$-th cluster, $\phi_{c,\ell}^r$ and $\theta_{c,\ell}^r$ are the angles of arrival (AoA) at the azimuthal and elevation planes while $\phi_{c,\ell}^t$ and $\theta_{c,\ell}^t$ are the angles of departure (AoD) in the azimuth and elevation planes. In addition,  $\textbf{a}_r(\phi_{c,\ell}^r,\theta_{c,\ell}^r)$ and $\textbf{a}_t(\phi_{c,\ell}^t,\theta_{c,\ell}^t)$ are the array response and steering vectors, respectively.

The channel at the $k$-th subcarrier is given by
\begin{equation}
\textbf{H}[k] = \sum_{n=0}^{N_c-1} \textsf{H}[n] e^{-j\frac{2\pi k}{K}n}   
\end{equation}
where $K$ is the number of subcarriers.

\subsection{Self-Interference Channel Model}
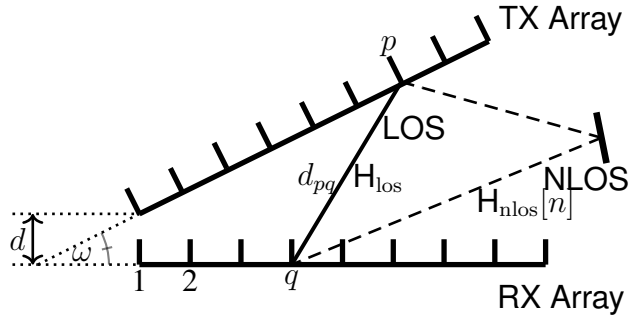
\begin{figure}[H]
\centering
\setlength\fheight{5cm}
\setlength\fwidth{15cm}

\usetikzlibrary{shapes.misc,shapes.geometric,shapes.symbols,positioning,shadings,automata}
\tikzstyle{XORgate} = [draw,circle]
\newcommand{\antena}{--++(3mm,0)--++(30:5mm)--++(-90:5mm)--++(150:5mm);}
\newcommand{\suma}{\Large$+$}

\definecolor{mycolor1}{rgb}{1.00000,0.00000,1.00000}%
\definecolor{orange}{rgb}{0.0,0.0,0.0}
\pgfdeclarelayer{background}
\pgfdeclarelayer{foreground}
\pgfsetlayers{background,main,foreground}
\usetikzlibrary{shapes,arrows}
\newcommand{\mx}[1]{\mathbf{\bm{#1}}} 
\newcommand{\vc}[1]{\mathbf{\bm{#1}}} 

\tikzstyle{sensor}=[draw, fill=yellow!30, text width=2em, 
    text centered, minimum height=2em, rounded corners]

\tikzstyle{chain}=[draw, fill=red!30, text width=2em, 
    text centered, minimum height=2em, rounded corners]
    
\tikzstyle{center1}=[draw=white,fill=white!20]   
    
\tikzstyle{ann} = [above, text width=5em]
\tikzstyle{node} = [sensor, text width=4em, fill=orange!50, 
    minimum height=19em, rounded corners]
   
\tikzstyle{circ} = [draw,circle,radius=0.5cm,fill=orange!50]
\def\blockdist{2.3}
\def\edgedist{2.5}
\def\antenna{%
    -- +(0mm,4.0mm) -- +(2.2mm,5.5mm) -- +(-2.2mm,5.5mm) -- +(0mm,4.0mm)
}

\usetikzlibrary{positioning}
\begin{tikzpicture}[thick,scale=0.675]
 \tkzDefPoint(0,0){A} \tkzDefPoint(1,0){B} \tkzDefPoint(1,.5){C} 
\tkzMarkAngle[fill= orange!40,size=1.4cm,opacity=.5](B,A,C)
\tkzLabelAngle[pos=0.9](B,A,C){$\omega$}  
\draw [-,dotted,line width=1pt] (0,0) to (2,0);                           
\draw [-,dotted,line width=1pt] (0,0) to (2,1);                           
\draw [-,orange,line width=2pt] (2,0) to (10,0); 
\draw [-,orange,line width=2pt] (2,0) to (2,0.5);
\draw [-,orange,line width=2pt] (3,0) to (3,0.5);
\draw [-,orange,line width=2pt] (4,0) to (4,0.5);
\draw [-,orange,line width=2pt] (5,0) to (5,0.5);
\draw [-,orange,line width=2pt] (6,0) to (6,0.5);
\draw [-,orange,line width=2pt] (7,0) to (7,0.5);
\draw [-,orange,line width=2pt] (8,0) to (8,0.5);
\draw [-,orange,line width=2pt] (9,0) to (9,0.5);
\draw [-,orange,line width=2pt] (10,0) to (10,0.5);
\node[align=right] at (10.3,-0.7){\sffamily{RX Array}};
\node[] at (2,-0.3){1};
\node[] at (3,-0.3){2};
\node[] at (5,-0.3){$q$};

\draw [-,orange,line width=2pt] (2,1) to (8.88,4.4); 
\draw [-,orange,line width=2pt] (2,1) to (1.75,1.5);
\draw [-,orange,line width=2pt] (2.86,1.4) to (2.61,1.9);
\draw [-,orange,line width=2pt] (3.72,1.9) to (3.47,2.4);
\draw [-,orange,line width=2pt] (4.58,2.3) to (4.33,2.8);
\draw [-,orange,line width=2pt] (5.44,2.7) to (5.19,3.2);
\draw [-,orange,line width=2pt] (6.3,3.1) to (6.05,3.6);
\draw [-,orange,line width=2pt] (7.16,3.6) to (6.91,4.1);
\node[] at (6.91,4.3){$p$};
\draw [-,orange,line width=2pt] (8.02,4) to (7.77,4.5);
\draw [-,orange,line width=2pt] (8.88,4.4) to (8.63,4.9);

\node[align=right] at (10.3,4.8){\sffamily{TX Array}};

\draw [-,line width=1.5pt] (5,0) to (7.16,3.6);
\node[] at (5.5,1.7){$d_{pq}$~};
\node[] at (6.7,1.7){\sffamily{$\textsf{H}_{\rm los}$}};
\node[] at (7.3,2.7){~\sffamily{LOS}};

\draw [-,black,line width=2.5pt] (11.2,2) to (11,3);
\node[] at (10.8,1.7){\sffamily{NLOS}};

\draw [-,dash pattern={on 5pt off 3pt on 0pt off 0pt} ,line width=1pt] (7.16,3.6) to (11.1,2.5);
\draw [-,dash pattern={on 5pt off 3pt on 0pt off 0pt} ,line width=1pt] (11.1,2.5) to (5,0);

\node[] at (9.5,1.2){~\sffamily{$\textsf{H}_{\rm nlos}[n]$}};

\draw [-,dotted,line width=1pt] (-0.5,0) to (0,0);                                                      
\draw [-,dotted,line width=1pt] (-0.5,1) to (2,1);                                                      
\draw [<->,line width=1pt] (-0.1,0) to (-0.1,1);

\node[align=left] at (-0.4,0.5){$d$};

\end{tikzpicture}
    \caption{Relative position of TX and RX arrays at BS. Given that the TX and RX arrays are collocated, the far-field assumption that the signal impinges on the antenna
    array as a planar wave does not hold. Instead, for FD transceivers, it is more suitable to assume that the signal impinges on the array as a spherical wave for the near-field LOS channel.}
     \label{position}
\end{figure}
As illustrated in Fig.~$\ref{position}$, the SI leakage at the BS is modeled by the channel matrix $\textsf{H}_s[n]$. Note that the SI channel is decomposed into line-of-sight (LOS) component modeled by $\textsf{H}_{\rm{los}}$ and non-line-of-sight (NLOS) leakage described by $\textsf{H}_{\rm{nlos}}[n]$. With larger delay spread, the channel $\textsf{H}_{\rm{nlos}}[n]$ is also frequency-selective such as the uplink and downlink channels defined by (\ref{channel}), while the channel $\textsf{H}_{\rm{los}}$ is static and depends on the geometry of the transceiver arrays. The LOS SI leakage matrix can be written as \cite{chal2}
\begin{equation}\label{eq2.2}
 [\textsf{H}_{\rm{los}}]_{pq} = \frac{1}{d_{pq}}e^{-j2\pi\frac{d_{pq}}{\lambda}}    
\end{equation}
where $d_{pq}$ is the distance between the $p$-th antenna in the TX array and $q$-th antenna in the RX array at BS. The aggregate SI $n$-th tap $\textsf{H}_s[n]$ can be obtained by
\begin{equation}\label{eq2.3}
\textsf{H}_s[n] = \underbrace{\sqrt{\frac{\kappa}{\kappa+1}}\textsf{H}_{\rm{los}}}_{\textsf{Near-Field}} + \underbrace{\sqrt{\frac{1}{\kappa+1}}\textsf{H}_{\rm{nlos}}[n]}_{\textsf{Far-Field}}    
\end{equation}
where $\kappa$ is the Rician factor.

\subsection{Antenna Array Model}
In this work, we propose the uniform rectangular array (URA) with $N \times M$ elements where $N$ and $M$ are the vertical and horizontal dimensions of the array/subarray, respectively. This model also encompasses special cases of array structure such as the uniform linear array (ULA) or uniform square planar array (USPA). The array response of the URA is given by
\begin{equation}
\begin{split}
\textbf{a}(\phi,\theta) =&
 \frac{1}{\sqrt{NM}}\left[1,\ldots,e^{j\frac{2\pi}{\lambda}(d_hp\sin\phi\sin\theta+d_vq\cos\theta)},\right.\\&\left.\ldots ,e^{j\frac{2\pi}{\lambda}(d_h(M-1)\sin\phi\sin\theta+d_v(N-1)\cos\theta)}   \right]^{T}
\end{split}
\end{equation}
where $\lambda$ is the signal wavelength, $d_h$ and $d_v$ are the antenna spacing in horizontal and vertical dimensions, respectively, $0 \leq p \leq M-1$, and $0 \leq q \leq N-1$ are the antennas indices in the 2D plane.

\subsection{Analog Beam Codebook}
Since 3D beamforming is assumed, we quantize the azimuth $\phi$ and elevation $\theta$ angles along with an oversampling factor $\rho$ as $\phi_m, m = 1,\ldots,M$ and $\theta_n, n = 1,\ldots,N$. The $m$-th element $\nu_{m,k,\ell}$ of azimuthal beam $\nu_{k,\ell}$ and the $n$-th element $\delta_{n,\ell}$ of elevation beam $\delta_\ell$ are given by
\begin{equation}
    \nu_{m,k,\ell} = \frac{1}{\sqrt{M}}\exp\left(-j\frac{2\pi}{\lambda}(m-1)d_h\sin\phi_k\sin\theta_\ell\right)
\end{equation}
\begin{equation}
\delta_{n,\ell} = \frac{1}{\sqrt{N}}\exp\left(-j\frac{2\pi}{\lambda}(n-1)d_v\cos\theta_\ell\right)  
\end{equation}
where $\phi_k$ and $\theta_\ell$ are the $k$-th and $\ell$-th element of $\phi$ and $\theta$, respectively. Thereby, the $(k,\ell)$ entry of the codebook $\omega_{k,\ell}$ supporting the 3D beamforming is given by the Kronecker product of the azimuthal and elevation array responses as 
\begin{equation}
\omega_{k,\ell} = \nu_{k,\ell} \otimes \delta_\ell.   
\end{equation}

\subsection{Fully-Connected Structure}
For this structure, each RF chain is connected to all the phase shifters of the antenna array. Although this structure achieves higher rate as it provides more DoF, it is not energy-efficient since a large amount of power is required for the connection between the RF chains and the phase shifters.
\subsection{Partially-Connected Structure}
For this structure, each RF chain is connected to a subarray of antennas which reduces the hardware complexity in the RF domain. Although fully-connected structure outperforms the partially-connected in terms of achievable rate, the latter structure is well advocated for energy-efficient systems. Note that the analog beamformer has the following structure

\begin{equation}
\textbf{F}_{\text{RF}} =
  \begin{pmatrix}
    \textbf{f}_1 & 0 & \dots & 0 \\
    0 & \textbf{f}_2 & \dots & 0 \\
    \vdots & \vdots & \ddots & \vdots \\
    0 & 0 & \dots & \textbf{f}_{N_{\text{RF}}}
  \end{pmatrix}.
\end{equation}

Each RF chain consists of a precoder $\textbf{f}_n,~n = 1\ldots N_{\text{RF}},$ which is a column vector of size $N_{\text{sub}} \times 1$ and $N_{\text{sub}}$ is the number of antennas of the subarray.

\subsection{Hardware Impairments}
Hardware imperfections, in particular the analog/RF front-end, present significant challenges in the SI suppression capabilities of FD transceivers. The primary imperfections are the transceiver phase and quantization noise and in-phase and quadrature (I/Q) imbalance as well as nonlinearities \cite{s5,s25} which also results in channel estimation errors.

High power amplifier (HPA) nonlinearities can substantially degrade system performance due to creating an irreducible error/outage floor and/or spectral efficiency saturation. The nonlinearities also create intermodulation products resulting in spectral regrowth and inter-carrier interference (ICI). 

Techniques to compensate the nonlinear effects include Bussgang Linearization Theory. Furthermore, related work proposed different HPA nonlinearities models such as Soft Envelope Limiter (SEL), Traveling Wave Tube Amplifier (TWTA) and Solid State Power Amplifier (SSPA) \cite{j1}. 

\begin{remark}
Nonlinearities and other imperfections in mmWave analog/RF hardware have significant impact on FD transceiver communication performance. Modeling transceiver hardware impairments is out of the scope of this work; however, we treat these impairments as additional sources of SI. For example, the aggregate SI power used in this work is around 80 dB. The near-far problem incurs about 20-40 dB of SI (depending on whether the UE is near the BS, at mid-range or at the cell edge) and the remaining SI comes from transceiver impairments.
\end{remark}

\section{Hybrid Beamforming Design}
In this section, we provide the framework for the design of hybrid beamformers for each algorithm. Specifically, we will decompose the full-digital beamformers ($\textbf{F}[k] = \textbf{F}_{\text{RF}}\textbf{F}_{\text{BB}}[k]$) into analog and digital parts that are jointly designed under some constraints. Since the analog precoder $\textbf{F}_{\text{RF}}$ of size ($N_{\text{T}}\times N_{\text{RF}}$) is implemented using the analog phase-shifters, it has the constant amplitude constraint, i.e., $|[\textbf{F}_{\text{RF}}]_{m,n}|^2 = \frac{1}{N_{\text{T}}}$. Further, we assume that the angle of the analog phase shifters are quantized to a finite set of possible values. With these assumptions, $[\textbf{F}_{\text{RF}}]_{m,n}=\frac{1}{N_{\text{T}}}e^{j\theta_{m,n}}$, where $\theta_{m,n}$ is the quantized angle. The total power is constrained by normalizing the digital precoder such that $\| \textbf{F}_{\text{RF}}\textbf{F}_{\text{BB}}[k]\|_F^2=N_s,~k=0,\ldots,K-1$, where $\textbf{F}_{\text{BB}}[k]$ is the $k$-th digital precoder of size ($N_{\text{RF}}\times N_s$), $N_{\text{RF}}$ is the number of RF chain and $N_s$ is the number of spatial streams. Note that the combiner is also subject to these constraints.

Next, we need to select the subcarrier (uplink or downlink) that will be used to design the analog beamformers. Based on the following criterion, we will search for the index $k^\star$ of the subcarrier with the highest energy as follows
\begin{equation}\label{index}
    k^\star = \argmax\limits_{k = 1,\ldots, K}\|\textbf{H}[k]\|_{\rm F}^2
\end{equation}
For the self-interference channel, we select the subcarrier with the lowest energy as follows
\begin{equation}\label{indexsi}
    k^\star = \argmin\limits_{k = 1,\ldots, K}\|\textbf{H}_s[k]\|_{\rm F}^2
\end{equation}
For the sake of notation, we drop the index $k^\star$ from the subcarrier and we just assume that $\textbf{G} =  \textbf{H}[k^\star]$ and $\textbf{G}_s =  \textbf{H}_s[k^\star]$. The detailed analysis for the beamforming designs of each algorithm is discussed in the following subsections. 

\subsection{Algorithm I: Downlink User Scheduling}
\begin{algorithm}[t]
 \caption{Downlink user scheduling}\label{algo1}
 \begin{algorithmic}[1]
 \renewcommand{\algorithmicrequire}{\textbf{Input:}}
 \renewcommand{\algorithmicensure}{\textbf{Output:}}
 \renewcommand{\algorithmiccomment}{$\triangleright~$}
 \REQUIRE $\mathcal{F}$,~$\mathcal{W}$,~$\textbf{H}_s[k]$,~$\textbf{H}_u[k]$, for $k=1,\ldots,K$, $N_u$ users.
 \STATE Apply Eq.~(\ref{index}) to get the subcarrier with the highest energy for uplink $\textbf{G}_u$.
  \STATE Construct the analog precoder of uplink UE by selecting the $N_{\text{RF}}$ beams that maximize $\|\textbf{G}_u \textbf{F}_{\text{RF,UE}} \|_F$.
  \STATE Check the rank deficiency of the precoded channel $\textbf{G}_u \textbf{F}_{\text{RF,UE}}$ to get the permissible spatial streams.
  \STATE Construct the digital precoder of uplink UE by applying the SVD on each precoded subcarrier $\textbf{H}_u[k] \textbf{F}_{\text{RF,UE}}$.
  \STATE Construct the analog precoder of BS by selecting the $N_{\text{RF}}$ beams that minimize $\|\textbf{G}_s \textbf{F}_{\text{RF,BS}}\|_F$.
  \STATE Schedule the downlink UE among the set of downlink users that maximize the energy of the precoded channel.
  \STATE Check the rank deficiency of the precoded channel $\textbf{G}_d \textbf{F}_{\text{RF,BS}}$ to get the permissible spatial streams.
    \STATE Construct the digital precoder of BS by applying the SVD on each precoded subcarrier $\textbf{H}_d[k] \textbf{F}_{\text{RF,BS}}$.
    \STATE Construct the analog combiner of BS by selecting the $N_{\text{RF}}$ beams that maximize the uplink {\sffamily{SINR}}.
    \STATE Construct the MMSE digital combiners of BS and downlink UE for all subcarriers.
 \RETURN $\bold{W}_{\text{BS}}[k]$,~$\bold{F}_{\text{BS}}[k]$,~$\bold{W}_{\text{UE}}[k]$,~$\bold{F}_{\text{UE}}[k],~k=0,\ldots,K-1$.
 \end{algorithmic} 
 \end{algorithm}
This algorithm jointly designs the analog and digital stages. Note that this algorithm starts by designing the beamformers for the downlink scenario before the uplink and exploits the users diversity to enhance the downlink rate. We also assume that the BS schedules only one user and allocates the resources with TDMA sharing. 
 
\subsection{Algorithm II: Best Downlink Precoding}
This algorithm is quite different from the previous one. In particular, Algorithm \ref{algo2} starts by designing the beamformers for the uplink scenario before the downlink. The analog and digital stages are jointly designed similarly to algorithm I. 
\begin{algorithm}[t]
 \caption{Best downlink precoding}\label{algo2}
 \begin{algorithmic}[1]
 \renewcommand{\algorithmicrequire}{\textbf{Input:}}
 \renewcommand{\algorithmicensure}{\textbf{Output:}}
 \renewcommand{\algorithmiccomment}{$\triangleright~$}
 \REQUIRE $\mathcal{F}$,~$\mathcal{W}$,~$\textbf{H}_s[k]$,~$\textbf{H}_u[k]$,~$\textbf{H}_d[k]$, for $k=1,\ldots,K$.
 \STATE Apply Eq.~(\ref{index}) to get the subcarrier with the highest energy for uplink $\textbf{G}_u$ and downlink $\textbf{G}_d$.
  \STATE Construct the analog precoder of BS in downlink by selecting the $N_{\text{RF}}$ beams that maximize $\|\textbf{G}_d \textbf{F}_{\text{RF,BS}} \|_F$.
  \STATE Check the rank deficiency of the precoded channel $\textbf{G}_d \textbf{F}_{\text{RF,BS}}$ to get the permissible spatial streams.
  \STATE Construct the digital precoder of BS by applying the SVD on each precoded subcarrier $\textbf{H}_d[k] \textbf{F}_{\text{RF,BS}}$.
  \STATE Construct the analog combiner of downlink UE selecting the $N_{\text{RF}}$ beams that maximize $\|\textbf{W}^*_{\text{RF,UE}}\textbf{G}_d \textbf{F}_{\text{RF,BS}} \|_F$.
   \STATE Construct the analog precoder of uplink UE by selecting the $N_{\text{RF}}$ beams that maximize $\|\textbf{G}_u \textbf{F}_{\text{RF,UE}} \|_F$. 
    \STATE Check the rank deficiency of the precoded channel $\textbf{G}_u \textbf{F}_{\text{RF,UE}}$ to get the permissible spatial streams.
  \STATE Construct the digital precoder of uplink UE by applying the SVD on each precoded subcarrier $\textbf{H}_u[k] \textbf{F}_{\text{RF,UE}}$.
    \STATE Construct the analog combiner of BS by selecting the $N_{\text{RF}}$ beams that maximize the uplink {\sffamily{SINR}}.
     \STATE Construct the MMSE digital combiners of BS and downlink UE for all subcarriers.
 \RETURN $\bold{W}_{\text{BS}}[k]$,~$\bold{F}_{\text{BS}}[k]$,~$\bold{W}_{\text{UE}}[k]$,~$\bold{F}_{\text{UE}}[k],~k=0,\ldots,K-1$.
 \end{algorithmic} 
 \end{algorithm}
 
 We observe that the beamformers for uplink are designed similarly to Algorithm I. The second difference is related to downlink scenario wherein a single downlink UE exists and we search for the best analog beam from the codebook, unlike the first algorithm in which the downlink precoder is first selected from the codebook to minimize the SI power and then we schedule the best downlink UE.
 
\subsection{Algorithm III: Max Effective Channel Energy}
In this part, we will provide a detailed analysis of the third algorithm for hybrid beamforming. Unlike the previous algorithms, the analog and digital stages are designed independently. The analog stage is based on maximizing the sum energy of the effective channel as
\begin{equation}\label{exhaustive}
 \left(\textbf{W}_{\text{RF}},~\textbf{F}_{\text{RF}}\right) =    \argmax\limits_{\textbf{w}_u\in \mathcal{W},~\textbf{f}_v \in \mathcal{F}} \sum_{u=1}^{N_{\text{RF},r}}\sum_{v=1}^{N_{\text{RF},t}} |\textbf{w}^*_u \textbf{G}_{vu}\textbf{f}_v|^2
\end{equation}
where $\textbf{G}_{vu}$ is the subchannel between the TX $v$-th and RX $u$-th RF chains, $\textbf{w}_u$ and $\textbf{f}_v$ are the combiner and precoder at $u$-th RX and $v$-th TX RF chains, respectively. $\mathcal{F}$ and $\mathcal{W}$ are the TX and RX codebooks, respectively. $\textbf{W}_{\text{RF}}$ and $\textbf{F}_{\text{RF}}$ are the precoder and combiner matrices of sizes $N_{\text{R}} \times N_{\text{RF},r}$ and $N_{\text{T}} \times N_{\text{RF},t}$, respectively. Note that this analog beamforming design is applicable for uplink and downlink phases. The detailed steps are illustrated by Algorithm \ref{algo3}.

\begin{algorithm}[b]
 \caption{Max effective channel energy}\label{algo3}
 \begin{algorithmic}[1]
 \renewcommand{\algorithmicrequire}{\textbf{Input:}}
 \renewcommand{\algorithmicensure}{\textbf{Output:}}
 \renewcommand{\algorithmiccomment}{$\triangleright~$}
 \REQUIRE $\mathcal{F}$,~$\mathcal{W}$,~$\textbf{H}_u[k]$,~$\textbf{H}_d[k]$, for $k=1,\ldots,K$.
 \STATE Apply Eq.~(\ref{index}) to get the subcarrier with the highest energy for uplink $\textbf{G}_u$ and downlink $\textbf{G}_d$.
  \STATE Construct the analog precoders and combiners for uplink and downlink by applying beam search across the codebooks to solve Eq.~(\ref{exhaustive}).
  \STATE Check the rank deficiency of the effective channel $\textbf{W}^*_{\text{RF}} \textbf{G} \textbf{F}_{\text{RF}} $
   to get the permissible spatial streams for uplink and downlink transmissions.
   \STATE Construct the digital precoders by applying the SVD on each effective subcarrier $\textbf{W}_{\text{RF}}^*\textbf{H}[k] \textbf{F}_{\text{RF}}$ for uplink and downlink scenarios.
  \STATE Construct the MMSE digital combiners of BS and downlink UE for all subcarriers.
 \RETURN $\bold{W}_{\text{BS}}[k]$,~$\bold{F}_{\text{BS}}[k]$,~$\bold{W}_{\text{UE}}[k]$,~$\bold{F}_{\text{UE}}[k],~k=0,\ldots,K-1$.
 \end{algorithmic} 
 \end{algorithm}
 
 Unlike Algorithms I and II, Algorithm \ref{algo3} is mainly based on maximizing the received power while the design disregards the minimization of the SI effective energy. We will show later by the results that this algorithm suffers from severe degradation incurred by the SI power.
 \begin{remark}
 In Algorithm III, the analog precoding and combining designs are performed by a joint search in the codebooks. A straightforward approach is to go through the exhaustive beam search. Although this approach provides optimal beamformers, the computational complexity is prohibitive as the number of operations grows exponentially with the size of codebook (number of RF chains and number of antennas). For this reason, we propose a suboptimal beam search to reduce the complexity and maintain an acceptable rate compared to the exhaustive approach. Next, we provide details and complexity comparisons between these two approaches.
 \end{remark}
 
\subsubsection{Exhaustive Beam Search}
This approach searches for the optimal precoders and combiners by considering all the combinations from the TX and RX beam codebooks. Although this approach is optimal, it is not recommended because it requires high complexity on the order $\mathcal{O}\left(N_{\text{sub},t}^{N_{\text{RF},t}}N_{\text{sub},r}^{N_{\text{RF},r}}\right)$.

\subsubsection{Suboptimal Beam Search}
This approach aims to reduce the size of TX and RX beam codebooks while keeping the best analog beams for each sides. For each TX RF chain, we collect exactly $N_{\text{RF},r}$ best RX beams. Then, for each RX RF chain, we repeat the same beam search  with all the TX RF chains to collect exactly $N_{\text{RF},t}$ for each RX subarray. The new TX and RX beam codebooks contain at most $N_{\text{RF},t} \times N_{\text{RF},r}$ beams since the same beam can be redundant for more than one combination between TX and RX RF chains. Note that the reduced TX and RX beam codebooks have the same number of analog beams, which is sufficiently smaller than the regular codebook size designed for the TX or RX subarray. Note that the complexity of this approach is $\mathcal{O}\left(N_{\text{RF},t}^{N_{\text{RF},r}}N_{\text{RF},r}^{N_{\text{RF},t}}\right)$.
\subsubsection{Example}
Let's provide an example to illustrate this concept. Assume the following system setting as $N_{\text{RF},t} = 2$, $N_{\text{RF},r} = 4$, $N_{\text{sub},t} = 16$ and $N_{\text{sub},r} = 32$. Applying beam search between each TX RF chain and all the RX RF chains yields 4 best RX beams out of 32. Vice-versa, each RX RF chain corresponds to 2 TX beams out of 16. The new codebook has at most 8 beams which significantly reduces the complexity compared to the exhaustive beam search approach. 
\section{Performance Analysis}
\subsection{Spectral Efficiency}
\subsubsection{Exact Analysis}
For uplink scenario, the spectral efficiency can be expressed as
\begin{equation}
\mathcal{I}_u({\scriptsize{\textsf{SNR}}}) = \frac{1}{K}\sum_{k=1}^K\log\det\left(\textbf{I} + \frac{{\scriptsize{\textsf{SNR}}}}{KN_s}\textbf{Q}_u[k]^{-1}\overline{\textbf{H}}_u[k] \overline{\textbf{H}}^*_u[k]\right)
\end{equation}
where $\textbf{Q}_u[k]$ is the SI plus noise autocovariance matrix of the $k$-th uplink subcarrier given by
\begin{equation}
\textbf{Q}_u[k] = {\textsf{INR}}  \textbf{W}^*_{\text{RF}}\textbf{H}_{s}[k]\textbf{F}_{\text{RF}} + \overline{\textbf{W}}^*[k]\overline{\textbf{W}}[k]
\end{equation}
where the combiner $\overline{\textbf{W}}[k] = \textbf{W}_{\text{RF}} \textbf{W}_{\text{BB}}[k]$, the equivalent channel $\overline{\textbf{H}}_u[k] = \textbf{W}^*_{\text{BB}}[k]\textbf{H}_{u,\text{eff}}[k]\textbf{F}_{\text{BB}}[k]$, the effective channel $\textbf{H}_{u,\text{eff}}[k] = \textbf{W}^*_{\text{RF}}\textbf{H}_{u}[k]\textbf{F}_{\text{RF}}$, and $\textsf{INR}$ stands for the Interference-to-Noise Ratio.

For the downlink scenario, the spectral efficiency is obtained by
\begin{equation}
\mathcal{I}_d({\scriptsize{\textsf{SNR}}}) = \frac{1}{K}\sum_{k=1}^K\log\det\left(\textbf{I} + \frac{{\scriptsize{\textsf{SNR}}}}{KN_s}\textbf{Q}_d[k]^{-1}\overline{\textbf{H}}_d[k] \overline{\textbf{H}}^*_d[k]\right)    
\end{equation}
where $\textbf{Q}_d[k]$ is the noise autocovariance matrix of the $k$-th downlink subcarrier given by
\begin{equation}
\textbf{Q}_d[k] = \overline{\textbf{W}}^*[k]\overline{\textbf{W}}[k].
\end{equation}

Note that $\textbf{Q}_u[k]$ and $\textbf{Q}_d[k]$ are both scaled by the noise power.
\subsubsection{Full-Digital Design}
For this design, we consider the SVD precoder $\textbf{F}_{\text{BB}}[k]$ applied at the $k$-th subcarrier for the most left $N_s$ streams and the MMSE combiner is applied on the precoded subcarrier $\textbf{H}[k]\textbf{F}_{\text{BB}}[k]$. Note that these steps are applied for uplink as well as for downlink scenario.

\subsubsection{Upper Bound}
For interference-free scenario, the optimal beamformers diagonalize the channel. By applying the SVD successively on all subcarriers, we retrieve the singular values associated to each subcarrier matrix. For each subcarrier, the singular values are listed in descending order and we will extract the first $N_s$ modes associated to the spatial streams. Equivalently, the upper bound is given by \cite{mimo}
\begin{equation}
\mathcal{I}({\scriptsize{\textsf{SNR}}}) =  \frac{1}{K}\sum_{k=0}^{K-1}\sum_{\ell=0}^{N_s-1}\log\left(1 + \frac{{\scriptsize{\textsf{SNR}}}}{KN_s} \sigma_{\ell}\left(\textbf{H}[k] \right)^2  \right)  
\end{equation}
where $\sigma_\ell(\textbf{H})$ is the $\ell$-th singular value of the channel matrix \textbf{H}. Note that the upper bound derivation follows the same rules for uplink as well as the downlink scenario.
\subsection{Outage Probability}
Once a transmission strategy is specified, the corresponding outage probability for rate $R$ (bit/s/Hz) is then \cite[Eq.~(4)]{lozano}
\begin{equation}
    P_{\textsf{out}}(\textsf{\scriptsize{SNR}}, R) = \mathbb{P}[\mathcal{I}(\textsf{\scriptsize{SNR}})<R].
\end{equation}

With convenient powerful channel codes, the probability of error when there is no outage is very small and hence the outage probability is an accurate approximation for the actual block error probability. As justified in the literature, modern radio systems such as UMTS and LTE operate at a target error probability. Therefore, the primary performance metric is the maximum rate\footnote[2]{In this work, we define the notion of rate with outage as the average data rate that is correctly received/decoded at the receiver which is equivalent to the throughput. In other standards in the literature, the rate with outage is assimilated with the transmit data rate. The only difference is if we consider rate with outage as the throughput, we account for the probability of bursts (outage) and we multiply by the term (1-$\epsilon$), while for the transmit data rate, the term (1-$\epsilon$) is not accounted anymore.}, at each {\textsf{\scriptsize{SNR}}}, such that this threshold is not overtaken, i.e., \cite[Eq.~(5)]{lozano}
\begin{equation}
R_\epsilon(\textsf{\scriptsize{SNR}}) = \max_{\zeta}\left\{ \zeta: P_{\textsf{out}}(\textsf{\scriptsize{SNR}}, \zeta) \leq \epsilon \right\}    
\end{equation}
where $\epsilon$ is the target.

\subsection{Energy Efficiency}
The energy efficiency, expressed in bit/s/Hz/Watt or bit/Joule/Hz, is defined as the ratio between the spectral efficiency and the total power consumption. It is expressed as \cite[Eq.~(37)]{alternating}
\begin{equation}
\mathcal{J}(\textsf{\scriptsize{SNR}}) = \frac{\mathcal{I}(\textsf{\scriptsize{SNR}})}{P_{\text{common}}+N_{\text{RF}}P_{\text{RF}}+N_{\text{T}}P_{\text{PA}} + N_{\text{PS}} P_{\text{PS}}}    
\end{equation}
where $N_{\text{RF}}$ is the number of RF chain, $P_{\text{common}}$ is the common power of the transmitter, $P_{\text{RF}}$ is the power of the RF chain, $P_{\text{PA}}$ is the power of PA, and $P_{\text{PS}}$ is the power of the phase shifter. Note that $N_{\text{PS}}$ is given by
\begin{equation}
N_{\text{PS}} = \left\{
        \begin{array}{ll}
            N_{\text{T}}N_{\text{RF}} & \quad \text{Fully-connected} \\
            N_{\text{T}} & \quad \text{Partially-connected}
        \end{array}
    \right. .
\end{equation}

\section{Numerical Results}
In this section, we present the numerical results of the reliability metrics following their discussion. We validate the accuracy of the analytical expressions with Monte Carlo simulations \footnote[3]{For all cases, $10^6$ realizations of the random variables were generated to perform the Monte Carlo simulation in MATLAB.}. Throughout this section, we will analyze the robustness of each algorithm against the interference and other system parameters in terms of rate, outage probability and energy efficiency. We start by constructing the channel as follows: The paths gains are independently drawn from a circular complex Gaussian distribution, all with the same variance. The AoA and AoD are random, with uniformly distributed mean cluster angle and angular spreads. Then, we introduce the raised cosine filter for pulse shaping to construct the channel at each tap. Unless otherwise stated, we summarize the simulation parameters in Table \ref{sim}.

\begin{table}[t]
\renewcommand{\arraystretch}{1.0}
\caption{System Parameters \cite{e2,anum,alternating}.}
\label{sim}
\centering
\begin{tabular}{rl}
\bfseries Parameter & \bfseries Value\\
\hline
Carrier frequency& 28 GHz\\
Bandwidth & 850 MHz\\
BS subarray size & 32 = 8$\times$4\\
UE subarray size& 4 = 2$\times$2 \\
Antenna separation & $\frac{\lambda}{2}$\\
Antenna correlation&None\\
Separation between BS arrays ($d$) & 2$\lambda$\\
Angle between BS arrays ($\omega$)& $\frac{\pi}{6}$\\
Hardware connections & Partial \\
Number of clusters ($C$) & 6\\
Number of rays per cluster ($L$)& 8\\
Angular spread& 20$^{\circ}$\\
Rician factor ($\kappa$) & 5 dB\\
Signal-to-Interference Ratio (SIR)& -15 dB\\
Number of spatial streams ($N_s$)& 2\\
Number of RF chains ($N_{\text{RF}}$) & 2\\
Number of subcarriers ($K$)& 128\\
Number of taps ($N_c$) & 4\\
Oversampling factor ($\rho$) &1\\
Power amplifier ($P_{\text{PA}}$) & 100 mW\\
Power of RF chain ($P_{\text{RF}}$)& 100 mW\\
Power of phase shifter ($P_{\text{PS}}$) & 10 mW\\
\end{tabular}
\end{table}

\subsection{A Primary Comparison}
\begin{figure}[H]
\centering
\setlength\fheight{5.5cm}
\setlength\fwidth{7.5cm}
\input{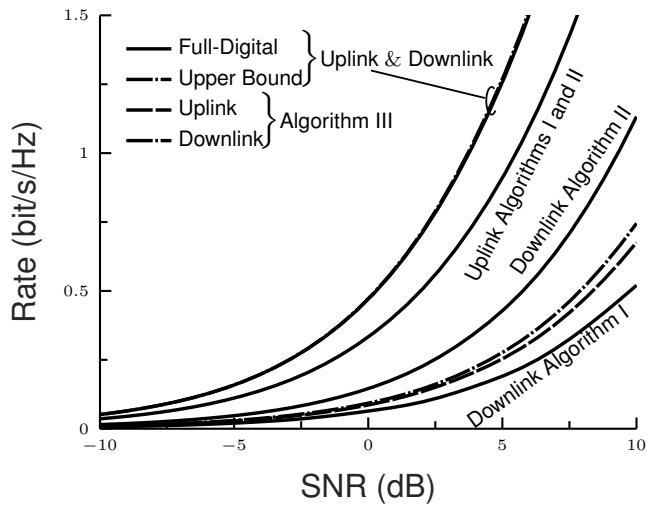}
    \caption{Rate performance results. Comparisons are made between the upper bound, full-digital and the three algorithms for hybrid beamforming design. The results are presented for uplink and downlink scenarios.}
    \label{pict1}
\end{figure}

Fig. \ref{pict1} illustrates the rate performance across a given range of {\sffamily{\scriptsize{SNR}}} for the three algorithms as well as full-digital and upper bound. Since the channels in uplink and downlink are symmetric ($\textbf{H}_u$ = $\textbf{H}_d^*$), the upper bound is similar for uplink and downlink transmissions. Similar to the upper bound, the uplink and downlink UEs achieve the same rate for full-digital design. We also observe that the full-digital design coincides with the upper bound performance as the interference is completely eliminated by the full-digital beamformers. Furthermore, the uplink rates for Algorithms I and II are quite similar since the beamforming for uplink is based on the same criterion (step 9 in Algorithms \ref{algo1} and \ref{algo2}). We further notice that Algorithm III offers lower uplink rate compared to Algorithms I and II, since the analog stage does not manage the interference power. For downlink scenario, Algorithm II offers the best achievable rate but now Algorithm III outperforms Algorithm I. In fact, the downlink rate for Algorithm I is highly dependent on the scheduled user and this rate loss is mainly explained by the poor channel of the downlink user and/or lack of user diversity. In the next discussion, we will show how the downlink rate for Algorithm I can be enhanced with user diversity.

\subsection{User Diversity and RF Chains}
\begin{figure}[H]
\centering
\setlength\fheight{5.5cm}
\setlength\fwidth{7.5cm}
\input{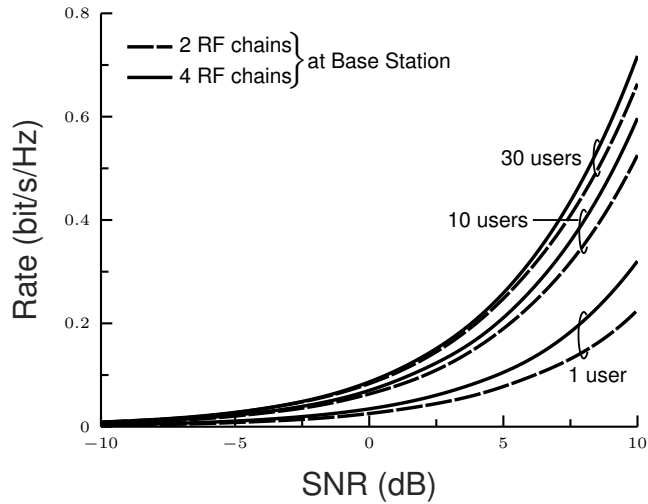}
    \caption{Rate performance: Results are evaluated for downlink scenario considering Algorithm I for different number of users. We further investigate the effect of the number of RF chains at the BS.}
    \label{pict2}
\end{figure}

Fig. \ref{pict2} provides the rate performance for downlink scenario (Algorithm I) considering various number of users and number of RF chains at the BS. First, we notice that the rate substantially improves with user diversity as in this case, the BS most likely schedules a user with strong channel. At a fixed {\sffamily{\scriptsize{SNR}}} of 10 dB and 2 RF chains, without user diversity (one user), the rate is around 0.2 bit/s/Hz while it jumps roughly to 0.5 and 0.7 bit/s/Hz for 10 and 30 users, respectively. This improvement is achieved by exploiting the user diversity gain. In addition, the performance can be further enhanced by increasing the number of RF chains. This result is expected since increasing the number of RF chains offers more DoF to limit the rate losses and provide the downlink UE with an acceptable rate. 

\subsection{Codebook and Signal-to-Interference Ratio (SIR)}
\begin{figure}[H]
\centering
\setlength\fheight{5.5cm}
\setlength\fwidth{7.5cm}
\input{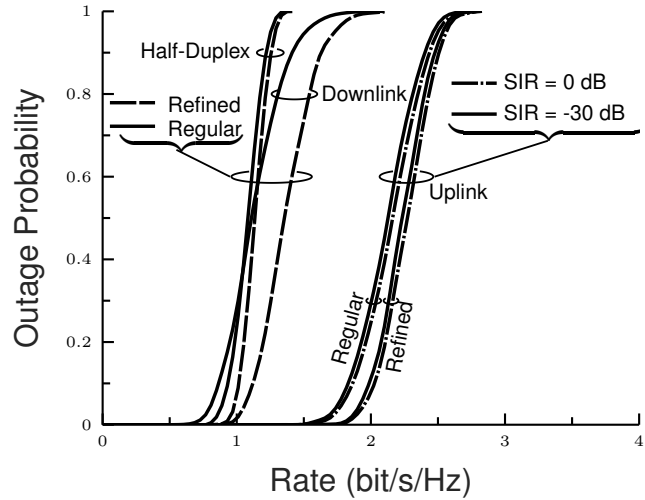}
    \caption{Outage probability performance: Comparisons are made between uplink, downlink FD and HD modes for Algorithm II. Evaluations are performed in terms of codebook oversampling factor and the {\sffamily{\scriptsize{SIR}}} levels. Regular and refined codebooks correspond to oversampling factors of 1 and 4, respectively. The {\sffamily{\scriptsize{SIR}}} levels of 0 and -30 dB can be interpreted as near and middle uplink users, respectively.}
    \label{pict3}
\end{figure}
Fig. \ref{pict3} illustrates the outage performance with respect to a given range of target rate requirements. In agreement with the results in Fig. \ref{pict1}, the uplink UE can support higher target rate compared to downlink UE and HD mode. With regular codebook, the gap between downlink UE and HD mode is relatively small and it gets slightly better up to 1.5 bit/s/Hz. This is explained by the fact that regular codebook is limited by the set of analog beams that cannot push up further the rate. With oversampling, the codebook becomes more refined as it offers more possible beam directions resulting in further enhancement to the received power and hence the rate. This improvement can be observed by comparing the HD and downlink UE modes. For example, the HD mode did not fully exploit the oversampling gain and it saturates roughly at 1.2 bit/s/Hz similar to regular codebook case. However, the margin between regular and refined cases is remarkable for downlink UE. Moreover, the uplink outage experiences further enhancement with codebook oversampling but this improvement margin is smaller than the downlink performance. These results are important to show that the performances are strictly governed by the codebook limitations. On the other side, the effect of the {\sffamily{\scriptsize{SIR}}} on the outage is roughly negligible and the near and middle users approximately saturate at the same threshold. This result can be interpreted by the robustness of beamforming of Algorithm II against the interference. This robustness is also measured by the pronounced margin gap with respect to the HD mode and downlink UE. 

\subsection{Beam Search and Duplex Modes}
\begin{figure}[H]
\centering
\setlength\fheight{5.5cm}
\setlength\fwidth{7.5cm}
\input{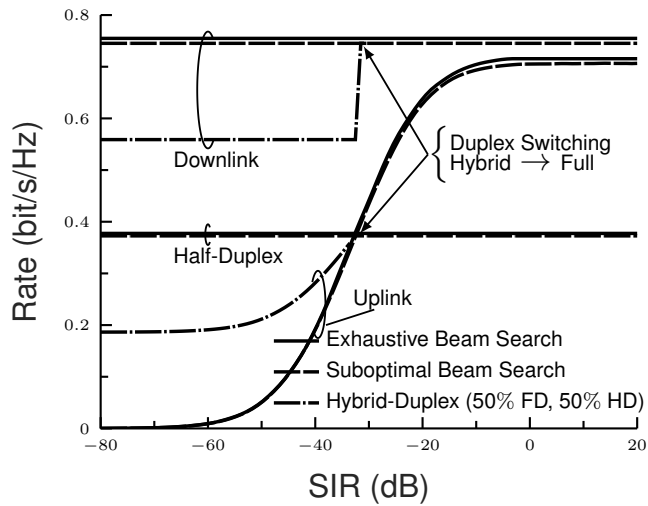}
    \caption{Rate performance results: Comparisons are made for Algorithm III relative to different scenarios. First comparison takes place between uplink, downlink in FD mode, HD and hybrid duplex. For the later duplex mode, the time duration is equally divided into FD and HD operations at the BS. On the other side, the exhaustive and suboptimal beam search approaches are evaluated and compared in terms of achievable rate. Note that the {\sffamily{\scriptsize{SIR}}} is changing by fixing the transmit power ({\sffamily{\scriptsize{SNR}}} = 10 dB) and varying the SI power.}
    \label{pict4}
\end{figure}

Fig. \ref{pict4} illustrates the variations of the achievable spectral efficiency with respect to the {\sffamily{\scriptsize{SIR}}} for Algorithm III. Since the downlink FD and HD modes are interference-free, the corresponding rates are constant since the {\sffamily{\scriptsize{SNR}}} is fixed at 10 dB. For a low SIR range from -80 to -40 dB which corresponds to a cell-edge user, the uplink rate is substantially degraded since Algorithm III disregards the interference cancellation and hence the uplink UE becomes completely exposed to the high SI power resulting in a practical null rate. Within the same range, the BS can operate in HD mode to avoid the severe interference and offers the uplink UE with relatively acceptable rate. However, this approach will also decrease the downlink rate from 0.77 to around 0.376 bit/s/Hz. 
A practical solution can be applied by introducing a new operating scheme called hybrid duplex to establish a tradeoff between the uplink and downlink rates. In this case, the uplink cell-edge user still achieves an acceptable rate around 0.19 bit/s/Hz instead of HD (0.376 bit/s/Hz) but this duplex mode offers better downlink rate roughly 0.58 bit/s/Hz. Although hybrid duplex improves the uplink cell-edge user at the expense of the downlink rate, the downlink UE still achieves better rate compared to HD mode. Starting from an {\sffamily{\scriptsize{SIR}}} of -30 dB and up to 20 dB (for middle and near users), it is recommended to switch from hybrid duplex operation to FD mode. These remarks lead to think about how to further improve the uplink and downlink rates for hybrid duplex mode. In other terms, we need to dedicate a careful attention on how to design two optimal switching points that have to be primarily adaptive to the {\sffamily{\scriptsize{SIR}}} level to maximize the uplink and downlink rates. The first switching occurs within the hybrid duplex mode, i.e. how to optimally allocate the time fractions for FD and HD, while the second switching occurs between hybrid and FD modes. For now, we defer the design of these optimal switching points as a future extension for this work. On the other side, we observe that the performances relative to exhaustive and suboptimal beam search approaches are relatively similar and the difference is quite negligible within the range of -20 to 20 dB of {\sffamily{\scriptsize{SIR}}}. Consequently, it is straightforward to adopt the suboptimal beam search method to reduce the complexity as it achieves relatively similar performance to exhaustive search.

\subsection{Hardware Connections}

\begin{figure}[H]
\centering
\setlength\fheight{5.5cm}
\setlength\fwidth{7.3cm}
\input{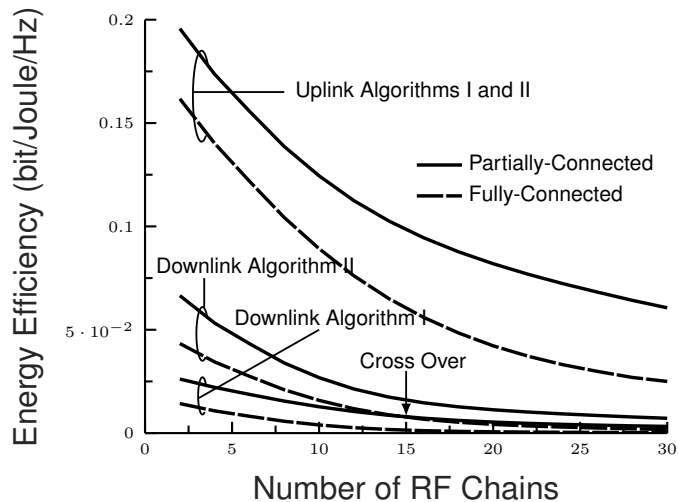}
    \caption{Energy efficiency performance: Results are evaluated for the three algorithms and for different numbers of RF chains at UEs and BS. Comparison is made between fully-connected and partially-connected structures.}
    \label{pict5}
\end{figure}

Fig. \ref{pict5} illustrates the variations of the energy efficiency for different number of RF chains at the UEs and the BS. In agreement with the conclusive summaries drawn for Fig. \ref{pict1}, the Algorithm II outperforms the other two algorithms not only in terms of rate but also in energy efficiency. We observe that all the performances are decreasing with the number of RF chains for partially and fully hardware connections. This observation shows that the spectral efficiency increases at a lower rate compared to the total power consumption which increases linearly with the number of RF chains. Moreover, the partially-connected is more power-efficient compared to the fully-connected structure since the latter requires a huge power consumption to support the full-connections established between each RF chain and all the antennas elements of the array. It is noteworthy to notice that there is a cross over between Algorithms I and II around 15 RF chains. Although Algorithm II achieves better downlink rate compared to Algorithm I, the latter implemented in partially-connected, outperforms the downlink Algorithm II implemented in fully-connected in terms of energy efficiency.

\subsection{Rate Gain/Loss}
In the sequel, we further investigate the effects of the interference on Algorithm III as a function of another metric by considering different {\scriptsize{\textsf{SIR}}} levels for uplink cell-edge, middle and near users. For this evaluation, we maintain the {\scriptsize{\textsf{SNR}}} fixed at 5 dB. To evaluate the efficiency of Algorithm III, we further define a new metric called the rate gain/loss as follows
 \begin{equation}
 \Gamma[\%] = \frac{ \mathcal{I}^{\text{FD}}(\scriptsize{\textsf{SNR}) } - \mathcal{I}^{\text{HD}}(\scriptsize{\textsf{SNR}) } }{ \mathcal{I}^{\text{HD}}(\scriptsize{\textsf{SNR}) } } \times 100   
 \end{equation}

In agreement with the conclusions drawn from Fig. \ref{pict4}, the uplink rate is completely deteriorated for low {\scriptsize{\textsf{SIR}}} range from -30 to -60 dB and the rate loss is pronounced as it lies within -26.02$\%$ and -99.49$\%$. These degradations can be alleviated by increasing the number of RF chains resulting in increasing the DoF to enhance the rate. With 4 RF chains, TABLE \ref{gain-loss} shows rate gain across all the {\scriptsize{\textsf{SIR}}} range and hence the BS can operate at FD mode and offers the uplink cell-edge user with an acceptable rate gain between 17.16$\%$ and 39.97$\%$.
 
 \begin{table}[H]
\renewcommand{\arraystretch}{1.0}
\caption{Rate Gain/Loss for Algorithm III}
\label{gain-loss}
\centering
\begin{tabular}{|l|c|c|c|c|c|}
\hline\hline
\multicolumn{6}{|c|}{2 RF chains at BS}\\\hline
{\scriptsize{\sffamily{SIR}}} (dB) & -20 & -30 & -40 & -50 & -60\\\hline
Uplink [\%] & 39.13 & -26.02 & -77.38 & -95.55 & -99.49\\\hline
Downlink [\%] & 100 & 100 & 100 & 100 & 100\\\hline
Sum [\%] & 69.56 & 36.99 & 11.30 & 2.22 & 0.25\\\hline
\multicolumn{6}{|c|}{4 RF chains at BS}\\ \hline

\scriptsize{\sffamily{SIR}} (dB) & -20 & -30 & -40 & -50 & -60\\\hline
Uplink [\%] & 72.5414 & 39.9745 & 21.7557 & 18.3583 & 17.1631\\\hline
Downlink [\%] & 100 & 100 & 100 & 100 & 100\\\hline
Sum [\%] & 86.2707 & 69.9873 & 60.8779 & 59.1792 & 58.5816\\\hline\hline
    \end{tabular}
 \end{table}

\textcolor{black}{\subsection{Algorithm II vs\ Conventional Approaches}
\begin{figure}[H]
\centering
\setlength\fheight{5.5cm}
\setlength\fwidth{7.5cm}
%
%
\begin{tikzpicture}

\begin{axis}[%
width=0.951\fwidth,
height=\fheight,
at={(0\fwidth,0\fheight)},
scale only axis,
xmin=-30,
xmax=10,
xlabel style={font=\color{white!15!black}},
xlabel={\textsf{SNR (dB)}},
ymin=0,
ymax=4,
ylabel style={font=\color{white!15!black}},
ylabel={\textsf{Sum Spectral Efficiency (bits/s/Hz)}},
axis background/.style={fill=white},
axis x line*=bottom,
axis y line*=left,
legend style={at={(0.03,0.97)}, anchor=north west, legend cell align=left, align=left, draw=white!15!black}
]


\node[right, align=left, rotate=0]
at (axis cs:-12.0,1.4) {$\Big\}$};

\draw [-,black,line width=.5pt] (-10.3,1.4) to (-7.2,.8);

\node[right, align=left, rotate=45]
at (axis cs:-1,.6) {\scriptsize{\textsf{Angle Search}} \cite{unconst}};

\node[right, align=left, rotate=45]
at (axis cs:-1,2.3) {\scriptsize{\textsf{Proposed}}};

\draw [-,black,line width=.5pt] (-1.5,11.9) to (4.82,11.1);

\node [draw=none,fill=none] at (rel axis cs: 0.25,0.35) {\shortstack[l]{
\ref{p4} \scriptsize{\textsf{Beam Steering}} \cite{unconst} \\
\ref{p5} \scriptsize{\textsf{SVD Approach}} \cite{unconst}}};

\addplot [color=black, line width=1.3pt]
  table[row sep=crcr]{%
-30	0.0803764794083419\\
-28	0.116855359146854\\
-26	0.174324741199914\\
-24	0.230639003941203\\
-22	0.30488571149646\\
-20	0.396045031135297\\
-18	0.502793604321244\\
-16	0.625092613491949\\
-14	0.769977476015561\\
-12	0.935212037515625\\
-10	1.11280159216962\\
-8	1.31596666233185\\
-6	1.53256958170774\\
-4	1.75475050647098\\
-2	1.99938825828047\\
0	2.25929119046123\\
2	2.52689775356596\\
4	2.80656301185449\\
6	3.09709686226314\\
8	3.38381352553049\\
10	3.6499459815923\\
};

\addplot [color=black, dash pattern={on 13pt off 1pt on 0pt off 0pt},line width=1.3pt]
  table[row sep=crcr]{%
-30	0.0046876481991519\\
-28	0.00774158575103014\\
-26	0.0138104617339865\\
-24	0.0222875205344128\\
-22	0.0346674689893138\\
-20	0.0539285672106126\\
-18	0.0818653007473587\\
-16	0.124537071211144\\
-14	0.183935066156674\\
-12	0.267703677792166\\
-10	0.38213419109283\\
-8	0.530316829372867\\
-6	0.714410236661411\\
-4	0.924379072757469\\
-2	1.17685610376282\\
0	1.45470966194027\\
2	1.76661621725749\\
4	2.10611106945867\\
6	2.47776871573872\\
8	2.85648422188796\\
10	3.25219532880948\\
};

\addplot [color=black, line width=1.3pt]
  table[row sep=crcr]{%
-30	0.00184889673659651\\
-28	0.00323352873029627\\
-26	0.00598675544908671\\
-24	0.0093681660286158\\
-22	0.0147637383452792\\
-20	0.0223603863099917\\
-18	0.0348992059334634\\
-16	0.0545645878771974\\
-14	0.0835161953773514\\
-12	0.127693387892608\\
-10	0.189522882904444\\
-8	0.271671874004016\\
-6	0.38510671881575\\
-4	0.524809766756966\\
-2	0.697241024575706\\
0	0.908764492049844\\
2	1.16875024418126\\
4	1.45674245238484\\
6	1.77428127708611\\
8	2.11626217682318\\
10	2.44972335034747\\
};

\addplot [color=black, line width=1.3pt]
  table[row sep=crcr]{%
-30	0.00688878398214706\\
-28	0.0114052072741971\\
-26	0.0202595026417321\\
-24	0.0316466682313386\\
-22	0.0489234914471151\\
-20	0.075211180223269\\
-18	0.113765492518648\\
-16	0.16759089697016\\
-14	0.243884691731737\\
-12	0.342751043707466\\
-10	0.468743340463777\\
-8	0.619976057796956\\
-6	0.797596276296859\\
-4	0.996969290896541\\
-2	1.23097163926575\\
0	1.48373392448681\\
2	1.76225115522788\\
4	2.08042956217154\\
6	2.40991752282425\\
8	2.74877393490321\\
10	3.08959846200004\\
};

\addplot [color=black]
  table[row sep=crcr]{%
-6.875	0.540096189432334\\
-6.88380119557527	0.537187268584315\\
-6.89279010539319	0.534492901073698\\
-6.901952208251	0.532017439536976\\
-6.91127270315994	0.529764882977323\\
-6.92073653325559	0.52773887030437\\
-6.93032841012163	0.525942674455704\\
-6.9400328384876	0.524379197109574\\
-6.94983414126086	0.523050963997353\\
-6.95971648485228	0.521960120823328\\
-6.96966390475473	0.521108429798403\\
-6.97966033133304	0.520497266793324\\
-6.98968961578383	0.520127619116018\\
-6.99973555622325	0.520000083916637\\
-7.00978192386039	0.520114867222889\\
-7.0198124892142	0.520471783607209\\
-7.02981104833157	0.521070256486308\\
-7.03976144896412	0.521909319052621\\
-7.04964761666144	0.522987615836141\\
-7.0594535807387	0.524303404894126\\
-7.06916350007663	0.525854560625131\\
-7.07876168871217	0.527638577202831\\
-7.08823264117847	0.52965257262407\\
-7.0975610575533	0.531893293364625\\
-7.10673186817546	0.534357119635126\\
-7.11573025798917	0.537040071228682\\
-7.12454169047713	0.539937813950727\\
-7.1331519311437	0.543045666620736\\
-7.14154707051009	0.546358608634469\\
-7.14971354658453	0.549871288074544\\
-7.15763816677113	0.553578030356237\\
-7.16530812918191	0.557472847394526\\
-7.17271104331774	0.561549447277596\\
-7.17983495008466	0.56580124443115\\
-7.18666834111329	0.570221370257131\\
-7.1932001773501	0.574802684229644\\
-7.19941990689054	0.579537785430177\\
-7.20531748202521	0.584419024503471\\
-7.2108833754715	0.58943851601473\\
-7.21610859576448	0.594588151188203\\
-7.22098470178229	0.59985961100657\\
-7.22550381638235	0.605244379649964\\
-7.22965863912661	0.610733758252912\\
-7.23344245807506	0.616318878956986\\
-7.23684916062858	0.621990719236454\\
-7.23987324340362	0.627740116473786\\
-7.24250982112266	0.633557782761472\\
-7.24475463450615	0.639434319906247\\
-7.24660405715324	0.645360234611474\\
-7.24805510140005	0.651325953813156\\
-7.2491054231461	0.657321840144822\\
-7.24975332564114	0.663338207506281\\
-7.24999776222615	0.669365336711107\\
-7.2498383380242	0.675393491187572\\
-7.24927531057832	0.681412932707664\\
-7.24830958943549	0.687413937118785\\
-7.24694273467726	0.693386810052701\\
-7.24517695439957	0.699321902586392\\
-7.2430151011456	0.70520962682947\\
-7.24046066729762	0.711040471413011\\
-7.23751777943524	0.716805016854773\\
-7.23419119166903	0.722493950775966\\
-7.23048627796053	0.728098082945004\\
-7.22640902344076	0.73360836012394\\
-7.22196601474155	0.739015880693583\\
-7.21716442935512	0.744311909033688\\
-7.21201202403911	0.749487889634981\\
-7.20651712228587	0.754535460920221\\
-7.20068860087627	0.759446468751973\\
-7.19453587553954	0.764212979605273\\
-7.18806888574264	0.768827293383899\\
-7.18129807863345	0.773281955859552\\
-7.17423439216382	0.777569770713845\\
-7.1668892374198	0.781683811163649\\
-7.1592744801875	0.785617431151021\\
-7.15140242178442	0.789364276079625\\
-7.1432857791872	0.792918293080315\\
-7.13493766448788	0.796273740789289\\
-7.12637156371186	0.799425198623016\\
-7.11760131503181	0.802367575534964\\
-7.10864108641268	0.805096118239964\\
-7.09950535272397	0.807606418892945\\
-7.09020887235616	0.809894422209618\\
-7.0807666633792	0.811956432017619\\
-7.07119397928141	0.813789117227515\\
-7.06150628432815	0.815389517214041\\
-7.05171922857994	0.816755046598864\\
-7.04184862261043	0.817883499427149\\
-7.03191041196515	0.818773052731189\\
-7.02192065140208	0.819422269475333\\
-7.01189547895594	0.819830100877451\\
-7.00185108986781	0.819995888103203\\
-6.99180371042247	0.819919363330352\\
-6.98176957173549	0.819600650181426\\
-6.97176488353255	0.819040263524005\\
-6.96180580796331	0.81823910863898\\
-6.95190843349218	0.817198479758103\\
-6.94208874890805	0.815920057973219\\
-6.93236261749508	0.814405908520524\\
-6.92274575140626	0.812658477444273\\
};

\end{axis}
\end{tikzpicture}%
    \caption{Sum spectral efficiency results: Comparisons are made between the proposed and conventional approaches. Note that the conventional techniques presented in \cite{unconst} are developed for machine to machine FD systems. In this work, we changed these techniques accordingly to support the proposed system model.}
    \label{pict21}
\end{figure}
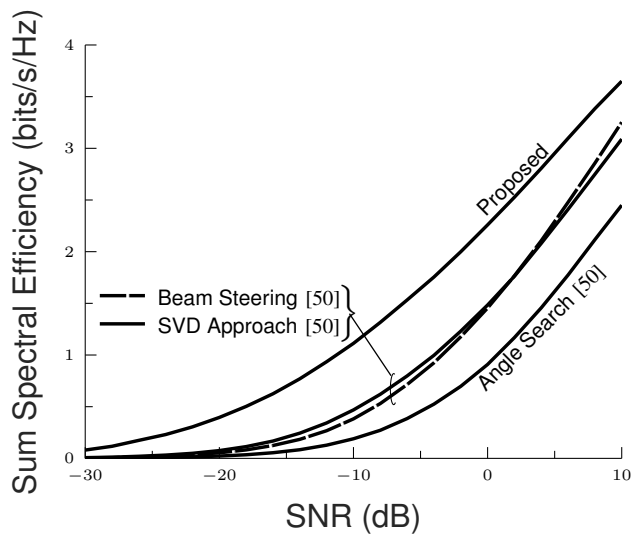 
Fig. \ref{pict21} compares proposed and conventional approaches implemented in analog-only architectures. We observe that conventional designs are very sensitive to the SI while the proposed design is more resilient to SI. The proposed algorithm achieves higher sum spectral efficiency around 12 bits/s/Hz at 10 dB of SNR whereas beam steering, SVD and angle search techniques achieve roughly 9, 8.7, and 7 bits/s/Hz.}

\section{Conclusion}
In this work, we proposed three algorithms of hybrid beamforming designs for a wideband FD system. The performances are measured in terms of spectral efficiency, outage probability and energy efficiency. We analyzed the effects of interference and other parameters on the system performance. We conclude that Algorithm II provides the best performance and outperforms the other two algorithms since it provides better sum rate. Algorithm I achieves a good uplink rate like Algorithm II but the downlink performance is highly dependent on the user diversity. We also showed that Algorithm III performance are relatively poor since the beamforming disregards the interference cancellation and instead maximizes the received power. However, these poor results can be enhanced by two approaches. Either by increasing the number of RF chains to offer more DoF or operating at hybrid duplex mode to mitigate the effects of the interference. Moreover, we discussed the impacts of the codebook and we ended up with performances that are prounouncedly limited by the codebook but further improvements can be achieved by oversampling. In addition, we illustrated the effects of the hardware connections of the energy efficiency and we validated the expected conclusion that the partially-connected is well advocated for systems consuming huge amount of power since it is more power-efficient than the fully-connected structure. Throughout this work, we observed that the performances are strictly limited by the codebook. Therefore designing more robust codebooks will certainly push up further the spectral efficiency and mitigate the effects of the interference. We are planning to consider this approach as a potential extension for this work. On the other side, instead of increasing the number of RF chains to offer more DoF, it is better to focalize more on designing the hybrid duplex mode  since more RF chains requires more power consumption. This duplex mode triggers us to think about how to design the optimal switching points which deserve a careful attention for our future work.

\bibliographystyle{IEEEtran}
\bibliography{main}
\end{document}